\documentclass[twocolumn,superscriptaddress,showpacs,preprintnumbers,amsmath,amssymb]{revtex4}
\usepackage{graphicx}
\usepackage{dcolumn}
\usepackage{bm}
\begin{document}
\title{Analysis by x-ray microtomography  of a granular packing undergoing compaction}
\author{Patrick Richard}
\email{Patrick.Richard@univ-rennes1.fr}
\affiliation{
G.M.C.M., UMR CNRS 6626, Universit\'e de Rennes I,
F-35042 Rennes Cedex, France}
\author{Pierre Philippe}
\affiliation{P.M.M.H., UMR CNRS 7636, ESPCI, 
10 rue Vaucquelin, F-75231 Paris cedex 05, France}
\author{Fabrice Barbe}
\affiliation{
L.M.R., UMR 6138, INSA Rouen,
Campus du Madrillet,
F-76801
Saint-\'Etienne-du-Rouvray Cedex, France}
\author{St\'ephane Bourl\`es}
\affiliation{
G.M.C.M., UMR CNRS 6626, Universit\'e de Rennes I,
F-35042 Rennes Cedex, France}
\author{Xavier Thibault}
\affiliation{ESRF, BP 220, F-38043 Grenoble Cedex, France}
\author{Daniel Bideau}
\affiliation{
G.M.C.M., UMR CNRS 6626, Universit\'e de Rennes I,
F-35042 Rennes Cedex, France}
\date{today}
\begin{abstract}
Several acquisitions of X-ray microtomography have been performed
on a beads packing while it compacts under vertical vibrations. An
image analysis allows to study the evolution of the packing
structure during its progressive densification. In particular, the
volume distribution of the pores reveals a large tail, compatible
to an exponential law, which slowly reduces as the system gets
more compact. This is quite consistent, for large pores, with the free
volume theory. These results are also in very good agreement with those
obtained by a previous numerical model of granular compaction.
\end{abstract}
\maketitle

Understanding the slow dynamics of out-of-equilibrium systems is
still an open and debated issue. Even though granular media are
not thermal systems, due to the insignificance of the thermal
energy $k_BT$ in comparison to the energy needed to move a
macroscopic grain, they are of great interest in this context.
Indeed, it has been suggested that the relaxation of a granular
medium under weak mechanical perturbations, such as shaking or
shearing, has a formal analogy with the slow dynamics of
out-of-equilibrium thermal systems \cite{Edwards,Nicodemi}. This
analogy is based on the idea that the geometry of the system is
more important than any other parameter as the driving energy or
the interaction between particles. Consequently, granular media
are often presented as ideal systems for studying the relaxation
towards equilibrium~\cite{Josserand2000}. Nevertheless, one may
keep in mind that, unlike the thermal energy, the mechanical
agitation of the granular material is neither stochastic non 
generally isotropic and thus leads to different spatial scales
of the energies \cite{Caglioti1999}.

One of the most common behavior of a shaken granular packing is
its slow compaction. This compaction is easily observed through
the progressive increase of the volume fraction and appears to be
a simple way to investigate the above analogy. The first
experiments on compaction have been realized in a thin tube giving
rise to a high lateral steric constraint in the packing
\cite{Chicago}. The evolution of the volume fraction reveals a
very slow dynamics and can be reasonably well fit by an empirical
law in the inverse of the logarithm of the number of shakes. More
recently, Philippe and Bideau carried out new experiments with a
reduced lateral confinement \cite{Philippe2002}. They showed that
the resulting compaction dynamics is consistent with a stretched
exponential,
also called Kholrausch-Williams-Watts
function, 
which
is the most usual description of the relaxation in a
out-of-equilibrium system such as a glass (see for instance
\cite{Phillips} and references therein).

Contrary to most previous studies on granular compaction in which
the quantities measured deal with the properties of, at least, a
part of the packing (as the mean packing fraction $\Phi$
\cite{Chicago,Philippe2002} or the vertical profile
\cite{Philippe2002}), the study discussed here benefits from x-ray
microtomography to characterize the microstructure of a packing
undergoing compaction. A previous work on x-ray microtomography
applied to granular materials has been carried out recently that
involves a structural study of a unique packing of about 2000
nearly monosized beads \cite{Seidler2000}. In this paper we focus
on the microstructural evolution of a compacting granular bed by
means of an original statistical analysis {and we relate 
this to the free volume theory}.

\begin{figure}
\includegraphics*[width=5cm]{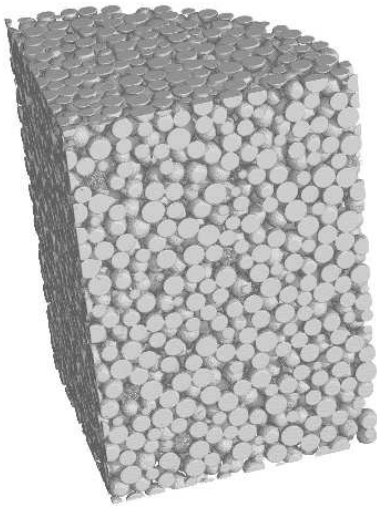}
\caption{\label{fig:joli} Part of the 3D reconstruction of the
initial packing ($\Phi\approx0.57$).}
\end{figure}

We use an experimental setup close to previous ones
\cite{Chicago,Philippe2002}: 200-400 $\mu$m diameter glass beads
are poured to about 80 mm height in a 8 mm inner-diameter glass
cylinder. The whole is vertically shaken by sinusoidal excitation
at a frequency of 70 Hz. The applied acceleration is measured by
an accelerometer and the intensity of the vibration is
characterized by $\Gamma$, the maximal applied acceleration
normalized by gravity ($g= 9.81$ ms$^{-2}$). The experiments are
performed as follows: starting from an initial loose reproducible
configuration ($\Phi\approx0.57$), a packing is vibrated for a
given number $N$ of oscillations with a fixed acceleration
$\Gamma$ and then analyzed by x-ray microtomography. Experimental
data acquisitions have been recorded at the ID19 beam line of the
ESRF (European Synchrotron Radiation Facilities). Synchrotron
radiation improves upon traditional x-ray techniques and offers
alternative methods of imaging. State-of-the-art microtomography provides
the three-dimensional mapping of the linear absorption coefficient
in the bulk of millimeter-sized samples with a spatial resolution
of the order of $1$ $\mu$m. A description of this experimental
method may be found in~\cite{Coles}. A monochromatic coherent beam
is used to get sample radiography for 1200 angular sample
positions ranging from $0^\circ$ to $180^\circ$. An x-ray energy of 51 keV
was selected to ensure a high enough signal to noise ratio. The
exposure time is 1 s by projection. The energy is selected
using a classical double monocrystal device. A filtered
backprojection algorithm~\cite{Herman1980,Natterer1986} is used to
compute the three-dimensional mapping of the linear absorption
coefficient in the sample. The light detector used is based on the
the fast read out low noise charge couple device
(FRELON CCD) camera
developed by the ESRF detector group ({1024 $\times$ 1024 elements, 14 bits
dynamic}).
A thin scintillation
layer deposited on glass converts x-rays to visible light. Light
optics magnify the image of the scintillator and project it onto
the CCD. With such a setup, the resolution is 9.81 $\mu m$ by
pixel.
Each reconstructed packing contains
around $15000$ grains. An example is reported in
Fig~\ref{fig:joli}.

Using an image processing software, the size and the location of
each particle are recovered. Then a complete set of information is
available to study the geometry of the packing. 
In the following
we note $\langle R\rangle$ and $\langle w\rangle$ the
mean radius and mean volume of a sphere, respectively.
\begin{figure*}[htb]
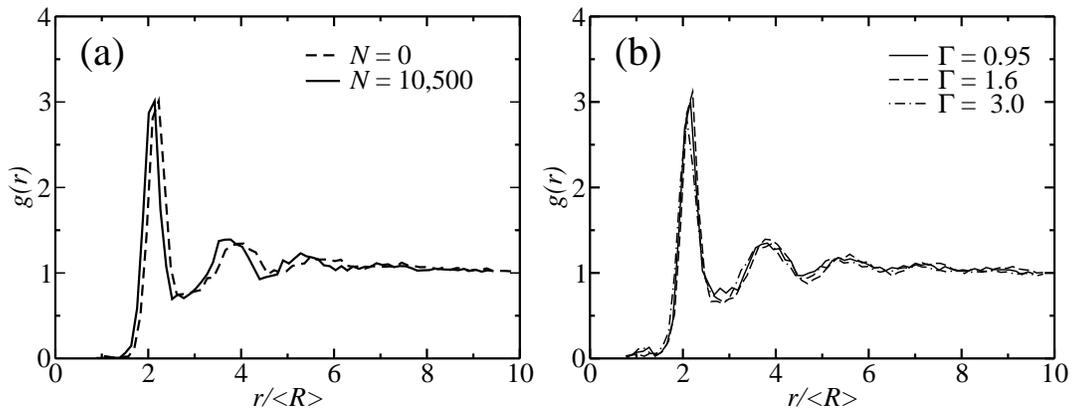

\includegraphics*[width=7cm]{fig2a.eps}
\includegraphics*[width=7cm]{fig2b.eps}
\caption{(a) Evolution of the pair correlation function for
$\Gamma=3.0$. (b) Steady-state pair correlation functions obtained
for $\Gamma=0.95$, $\Gamma=1.6$ and $\Gamma=3.0$.} \label{fig:gr}
\end{figure*}

The first characteristics reported are the classical pair
correlation function $g(r)$. Figure~\ref{fig:gr}(a) represents the
variation of the pair correlation function before and after more
than $10^{4}$ excitations of intensity $\Gamma=3.0$. Note that the
initial packing fraction, measured with the image processing
software, is equal to $0.57$, whereas the steady-state packing
fraction, obtained after a large number of excitations, is
approximately $0.62$. It appears that the different peaks of
$g(r)$ slowly move with the number of taps. In particular the
second peak slightly progresses from $r=4\langle R\rangle$ to
$r=2\sqrt{3}\langle R\rangle$, which is characteristic of
developing structure.
{Besides,  one can also remark that unlike  
Seidler et al.~\cite{Seidler2000}, we do not observe a double
 peak around $r = 4 \langle R \rangle$. 
This is due to large size distribution of the grains in
our experimental work (in~\cite{Seidler2000} 
it is significantly weaker.)}
Nevertheless the evolutions observed for $g(r)$ are not
significant. It is even more striking if one compares $g(r)$ for
steady-states corresponding to different intensities of excitation
as reported in Fig~\ref{fig:gr}(b). Here the steady-state pair
correlation functions obtained for $\Gamma=0.95$, $1.6$ and
$3.0$ have been plotted; their packing fractions are,
respectively,  $0.617$, $0.639$, and $0.624$. These correlation
functions do not differ significantly. It means that the pair
correlation function is not a suitable tool to study structural
changes during granular compaction. Compaction is then a process
more subtle than a simple decrease of the mean distance between
grains. Since this function is averaged over all the angles of
orientation, angular reorganizations cannot be seen. In contrast
to the numerical work of Rosato and Yacoub~\cite{Rosato2000}, we
do not observe the appearance of secondary sharp peaks in the pair
correlation function. This may be due to the difference in terms
of vibration intensity $\Gamma$
(up to $3$ in this work, $6.4$ in~\cite{Rosato2000})
 and to the small size of the vessel
they used, which can induce strong wall effects as
crystallization. 
{Note that by using a local bond order parameter~\cite{Q6}
we do not observe crystallization due to the
walls.
This is not surprising since
the size distribution of our grains is rather large. }
\begin{figure*}
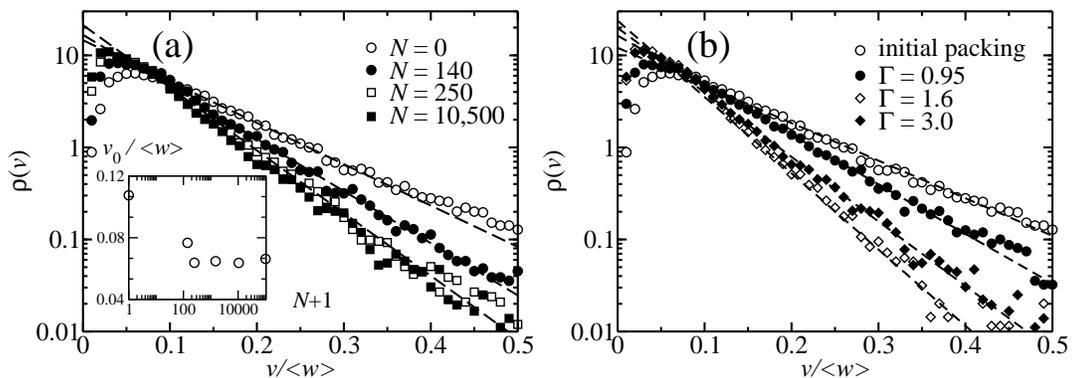

\includegraphics*[width=7cm]{fig3a.eps}
\includegraphics*[width=7cm]{fig3b.eps}
\caption{(a) Evolution of the volume distribution of the pores
during compaction with $\Gamma=3.0$. Inset of (a): decrease of
$v_0/\langle w\rangle$ with the number of oscillations $N$ and for
$\Gamma=3.0$. (b) Volume distributions of the pores for the
initial packing and for three different steady-state packings obtained
for $\Gamma=0.95$, $1.6$ and $3.0$
(b).}\label{fig:pores}
\end{figure*}

\begin{figure}
\includegraphics*[width=7cm]{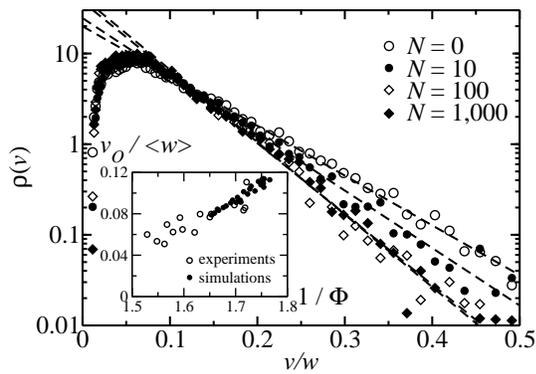}
\caption{Volume distributions of the pores obtained with the
numerical model of~\cite{Philippe2001} at different stages of a
packing undergoing compaction with the dilation parameter
$\epsilon=0.1$. Here $w$ is the volume of the
particles. Inset : plot of $v_0 / \langle w \rangle$ as a
function of $1/\Phi$ for all experiments ($\circ$) and 
a fully representative set of simulations
($\bullet$).}\label{fig:simu}
\end{figure}

Further evidence of a transformation in the packing microstructure
is provided by the study of the size of the interstitial voids.
{This can also be an experimental test for the
free volume theory~\cite{Boutreux1997,Caglioti1999} that
postulates an exponential decay for the distribution of these voids
$\rho(v)\propto \exp\left(-v/v_0\right)$.}
Following previous works~\cite{Richard1999,Philippe2001} we define
a pore size via the {Vorono\"\i} tessellation. A {Vorono\"\i}
polyhedron around a sphere is the region of space in which all
points are closer to this sphere than to any other sphere in the
packing. The {Vorono\"\i} network that is the whole collection of
the vertices and of the edges of the polyhedra maps the pore
space. Each vertex of this tessellation is equidistant to four
neighboring spheres and therefore defines the center of a pore.
The volume of a pore is then the size of a virtual sphere centered
on the vertex and in contact with the four neighboring spheres.
The volume of this void-sphere partially reflects the volume of
the whole void situated inside the tetrahedron formed by
the centers of four neighboring spheres. Since in our experiments
the spheres are not perfectly monosized, we have adapted this
method to a polydisperse packing replacing the {Vorono\"\i}
tessellation by the so-called navigation map~\cite{Richard2001}.
It is then possible to compute the size distribution of the pores
for each packing. Following Philippe and Bideau
\cite{Philippe2001} the volume of the pores $v$ is normalized by
the mean volume of a grain $\langle w\rangle$.
Figure~\ref{fig:pores}a presents the evolution of the volume
distribution of the pores for $\Gamma=3.0$ at different stages of
the compaction. First of all, we observe that unlike the pair
correlation function, $\rho(v/\langle w\rangle)$ drastically
changes with the number of excitations. Indeed an exponential
decay law is found for the distribution of the voids, at least for
the pores larger than octahedral pores
($v>(\sqrt2-1)^3\approx0.0711$). The exponential shape persists
during the compaction of the packing yet with a reduction of the
tail, i.e. with a decrease of the characteristic volume $v_0$. A
typical evolution of $v_0$ is shown in the inset of
Fig~\ref{fig:pores}(a). As reported in Fig~\ref{fig:pores}(b), the
final exponential decays of the steady-state distribution,
obtained after a sufficient number of oscillations, are strongly
dependent on the intensity $\Gamma$ of the vibrations. By
contrast, the part of the distribution corresponding to the
smallest pores is less $\Gamma$ dependent. The remarkable point is
that these results are very close to those obtained by numerical
simulation by Philippe and Bideau~\cite{Philippe2001}. Indeed, as
is illustrated in Fig.~\ref{fig:simu}, they also found an
exponential decay for $\rho(v/\langle w\rangle)$ which slightly
reduces as the system compacts. Their simulation is solely based
on the geometric constraints in a three-dimensional packing of
identical hard spheres. Each tap is decomposed into a vertical
dilation of intensity $\epsilon$ ($z\rightarrow z[1+\epsilon]$)
followed by a nonsequential deposition. This last stage is ruled
by gravity (the individual motions of the particles are
preferentially oriented along the vertical) and by the condition
of non interpenetration between particles. When the gravitational
energy of the whole packing gets nearly stabilized, the deposition
is stopped and another tap is implemented. More details can be
found in~\cite{Philippe2001}. The agreement between this
simulation and our experiments emphasizes the fundamental
importance of the geometry in granular compaction. The steric
constraint between the grains seems to rule the local
rearrangements of grains allowed by the shaking energy.
{One can observe that the small pore distributions
for the numerical simulations and the experiments
are quite different. We explain this point by the fact that contrary to our
experiments the grains used for the numerical simulations are perfectly
monosized and moreover they are never in real contact
(see~\cite{Philippe2001}.) 
Indeed the grain size distribution decreases the mean volume
of the pores and 
therefore the number of small pores is larger
in the experiments.
We can  also observe that
the existence of an exponential tail does not seem 
to be sensitive to the polydispersity. }
{However, we should point out that contrary to our experiments
and to previous work~\cite{Philippe2002}, significant
packing fraction gradients are present in the simulations.
They are probably consequences of the 
$z$-dependent dilation used in the model.}
{In order to quantify the importance of the tail in the pore
size distribution we also have plotted on the inset of
Fig.~\ref{fig:simu}, $v_0/\langle w \rangle$ as a function of
$1/\Phi$ for experiments and numerical simulations,
for any available number of taps and any shaking acceleration.
We observe a linear correlation between the 
two quantities which is in agreement with 
the free volume theory~\cite{Boutreux1997,Caglioti1999}.
This curve also shows that the packing fraction $\Phi$
and the characteristic volume $v_0$ vary in the same way
during the compaction.}\\
As demonstrated here, the use of X-ray microtomography to study
granular compaction provides much information on the
microstructure of a granular packing undergoing compaction. This
further opens several perspectives to this work. First of all, our
analysis gives an experimental measure of the free volume and can
be used as an experimental test of the free volume theory for
granular compaction (\cite{Boutreux1997,Caglioti1999}). Then, a
quantitative comparison between these microtomography experiments
and the numerical simulations described above can be done. In
particular, we plan to correlate the experimental excitation
parameter $\Gamma$ with $\epsilon$, the numerical one. For that
purpose we will use the measure of the characteristic volume of a
pore $v_0$ and assume this quantity provides the link between
$\Gamma$ and $\epsilon$. We also plan to study the hysteresis on
the steady-states and the microstructure of a granular material
during the compaction along the well-known reversible and
irreversible
branches~\cite{Nowak1997}.\\

In this article, we have reported experimental results on the
development of the microstructure in a granular packing submitted
to sinusoidal vibrations. It has been shown that, even at rather
high acceleration ($\Gamma=3$), no significant changes are
observed in the pair correlation function. The structural
signature of compaction is then more subtle than an average
decrease of the mean distance between grains. The reorganization
of the grains can be advantageously analyzed through the
distribution of the pores volumes. It has been found on a
consistent way with numerical simulations that compaction is
mainly due to a decrease of the number of the largest pores. The
volumes of these pores are statistically distributed along a broad
exponential tail which progressively reduces while the structure
gets more compact.
\ \\
\ \\
We are grateful to G. Vigier who is at the origin of this work. We
acknowledge the European Synchrotron Radiation Facility (ESRF) for
the use of their facilities, hospitality and
financial help. We are indebted to J. T. Jenkins and G. Le Ca\"er
for a careful reading of the manuscript. Two of us (P.R. and P.P.)
thank C. Menissez and S. Petitdidier for hospitality.

\end{document}